# UNCONDITIONALLY SECURE COMPUTERS, ALGORITHMS AND HARDWARE, SUCH AS MEMORIES, PROCESSORS, KEYBOARDS, FLASH AND HARD DRIVES[α]

LASZLO B. KISH[(1)] AND OLIVIER SAIDI[(2)]

*(1)Department of Electrical and Computer Engineering, Texas A&M University, College Station, TX 77843-3128, USA*

*(2) CRT Capital Holdings, LLC, 262 Harbor Drive, Stamford, CT 06902Y, USA*

*Version 3, April 23, 2008*

In the case of the need of extraordinary security, Kirchhoff-loop-Johnson-(like)-noise ciphers can easily be integrated on existing types of digital chips in order to provide secure data communication between hardware processors, memory chips, hard disks and other units within a computer or other data processor system. The secure key exchange can take place at the very first run and the system can renew the key later at random times with an authenticated fashion to prohibit man-in-the-middle attack. The key can be stored in flash memories within the communicating chip units at hidden random addresses among other random bits that are continuously generated by the secure line but are never actually used. Thus, even if the system is disassembled, and the eavesdropper can have direct access to the communication lines between the units, or even if she is trying to use a man-in-the-middle attack, no information can be extracted. The only way to break the code is to learn the chip structure, to understand the machine code program and to read out the information during running by accessing the proper internal ports of the working chips. However such an attack needs extraordinary resources and even that can be prohibited by a password lockout. The unconditional security of commercial algorithms against piracy can be provided in a similar way.

When a computer is lost, even if it is password protected, after disassembling it, the information stored in it will be accessible. Similarly, if a flash memory and portable hard drive backups are insecure. To defend against these problems data can be encrypted by software tools. However, these software encryption keys have limited security; they maybe eavesdropped, guessed or broken, if they are not sufficiently long.

The solution described in this short note offers a much higher level of data security by providing unconditionally secure data communication between the hardware elements of interest. Quantum encryption is an example where, with the help of the laws of physics, secure data communication is achieved [1,2]. However, those instruments are bulky and the price of a pair of communicators is around $100k which make them insufficient for this task. Recently, a statistical physics based circuitry, the *Kirchhoff-loop-Johnson-(like)-noise* (*KLJN*) cipher was proposed as an electronic competitor of quantum communicators [3,4], see Figure 1. To detect the eavesdropper, the KLJN system does *not* need to build a statistics (like quantum communicators do) and even the communication of a single bit is secure [2,3]. This is the only communicator which is

---

naturally secure against the man-in-the-middle attack [5]. The fully protected theoretical KLJN scheme with unconditional security against passive and active eavesdropping. The secret key is generated and shared by randomly choosing and connecting one of the resistors by Alice and Bob and then measuring the total loop resistance with their thermal noise or a properly enhanced noise [3]. Eve may actively try to extract information by injecting a current at the middle, see Figure 1. However, constantly monitoring, broadcasting and comparing the current and voltage values at the two ends uncover the eavesdropping and she can extract no bit without setting on the alarm [3]. If Eve breaks the line and executes a man-in-the-middle attack, that turns out to be an even weaker type of attack due to the statistical independence of generators at the two sides [5]. In the secure KLJN line communicators the public channel is protected by broadcasting the information in the public channel.

A system of communicator pairs can be expanded into an unconditionally secure data communication network [6]. Recently, a laboratory KLJN model was realized (see Figure 2) and tested via a model line and parameters up to 2000 km range were obtained. The prototype was experimentally tested with all the known/proposed breaking methods and the (raw-bit) security of the KLJN test model was superior to theoretical (raw-bit) estimations of quantum security. The KLJN prototype pair was placed on two computer cards (see Figure 2) however, simpler versions could have been simply integrated on a chip.

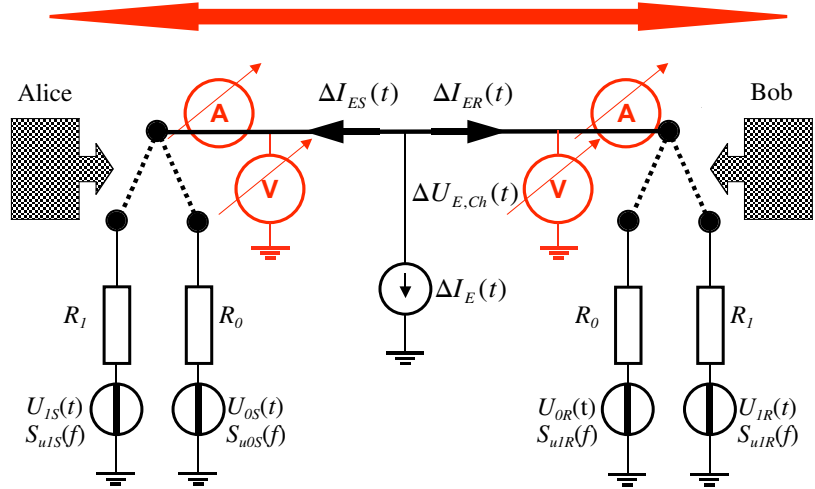

**Figure 1.** The fully protected theoretical KLJN scheme with unconditional security against passive and active eavesdropping. The secret key is generated and shared by randomly choosing and connecting one of the resistors by Alice and Bob and then measuring the total loop resistance with their thermal noise or a properly enhanced noise [3]. Eve may actively try to extract information by injecting a current at the middle. However, constantly monitoring, broadcasting and comparing the current and voltage values at the two ends uncover the eavesdropping and she can extract no bit without setting on the alarm [3]. If Eve breaks the line and executes a man-in-the-middle attack, that turns out to be an even weaker type of attack due to the statistical independence of generators at the two sides [5]. In the secure hardware, after the first key is made, the public channel should carry data encrypted with the shared key together with a hidden changing signature for authentication.

In the case of the need of extraordinary hardware, algorithm or data security [9], KLJN ciphers can easily be integrated on existing types of digital chips in order to provide secure data communication between hardware processors, memory chips, hard disks and other units within a computer or other data processor system [10]. Therefore, we propose that each secured computer hardware part has a sufficient number of integrated KLJN communicator units. In the simplest case, each unit separately communicates and exchanges a key with all the others. In this case, the required number of KLJN communicators in each hardware element is *N*-1 when *N* elements (chips) are connected securely, see Figure 3. Alternatively, if the number of units is too high, the secure network solution can be applied [6] which requires at least 2 KLJN communicators in each unit, but requires a more sophisticated protocol to run [6].

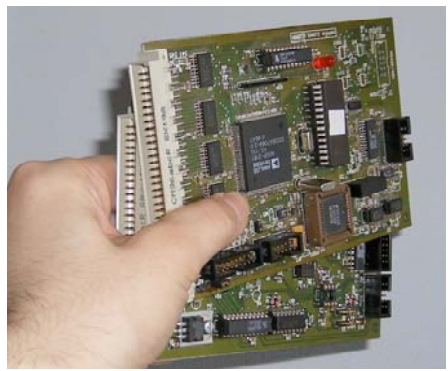

**Figure 2.** The realized KLJN network unit with a pair of KLJN communicators [7]. The computer chip version can be much simpler because of the very short cable distances in the computer providing idealistic conditions to approach the theoretical limits of performance.

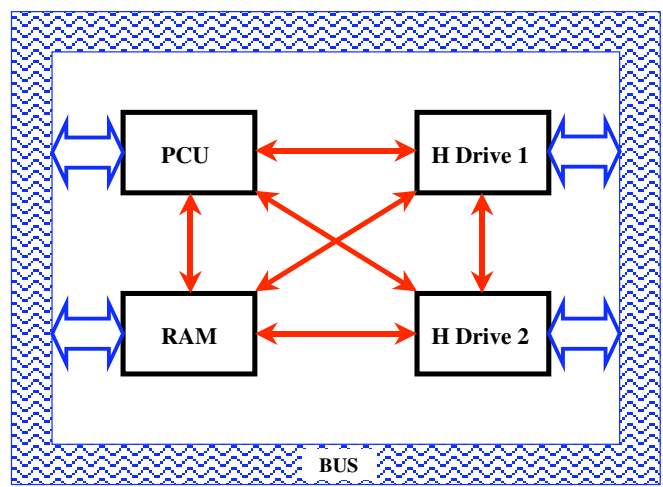

**Figure 3.** Example for securing a subsystem of a PCU, a RAM and two hard drives. Solid arrows: KLJN connections; Block arrows: classical data bus connections. Each unit has 3 KLJN communicators integrated on their chip. The required number of KLJN units/chip is N-1 when N chip gets connected securely.

The secure key exchange can take place at the very first run and the system can renew the key later at random times with an authenticated fashion to prohibit man-in-the-middle attack. In the secure hardware, after the first key is made, the public channel should carry data encrypted with the shared key together with a hidden changing signature for authentication. Thus after the first few milliseconds of the very first run, the communication between chips becomes unconditionally secure. The key can be stored in flash memories within the communicating chip units at hidden random addresses among other random bits that are continuously generated by the secure line but are never actually used. Thus, even if the system is disassembled, and the eavesdropper can have direct access to the communication lines between the units, or even if she is trying to use a man-in-the-middle attack, no information can be extracted. The only way to break the code is to learn the chip structure, to understand the machine code program and to read out the information during running by accessing the proper internal ports of the working chips. However such an attack needs extraordinary resources.

### Acknowledgment

In 2006, during a discussion about the potentials of KLJN ciphers, Olivier Saidi came up with a closely related idea to use KLJN to secure algorithm in digital games however, unfortunately, that time the issue was not worked out further.